\documentclass[12pt]{article}

\begin{document}

\title{{\Large \textbf{On tilted perfect fluid Bianchi type VI}}$_0$ {\Large 
\textbf{self-similar models}}}
\author{Pantelis S. Apostolopoulos\thanks{%
Permanent Address: University of Athens, Department of Physics, Section of
Astrophysics-Astronomy-Mechanics, Panepistemiopolis Zografos 15783, Athens,
Greece; \emph{e-mail}: papost@phys.uoa.gr.} \\
{\small \textit{University of Athens, Department of Physics, Nuclear and
Particle Physics Section,}}\\
{\small \textit{Panepistemiopolis Zografos 157 71, Athens, Greece}}}
\maketitle

\begin{abstract}
We show that the tilted perfect fluid Bianchi VI$_0$ family of self-similar
models found by Rosquist and Jantzen [K. Rosquist and R. T. Jantzen, \emph{%
Exact power law solutions of the Einstein equations}, 1985 Phys. Lett. 
\textbf{107}A 29-32] is the most general class of tilted self-similar models
but the state parameter $\gamma $ lies in the interval $\left( \frac
65,\frac 32\right) $. The model has a four dimensional stable manifold
indicating the possibility that it may be future attractor, at least for the
subclass of tilted Bianchi VI$_0$ models satisfying $n_\alpha ^\alpha =0$ in
which it belongs. In addition the angle of tilt is asymptotically
significant at late times suggesting that for the above subclasses of models
the tilt is asymptotically extreme.
\end{abstract}

\newline
In cosmological context \emph{transitively self-similar} Bianchi models
(spatially homogeneous models admitting a four dimensional group of
homotheties acting simply transitively on the spacetime manifold) keep a
privilege position among other high symmetric models (i.e. spacetime
homogeneous models which admit a simply transitive group of isometries or
solutions with special geometric or algebraic properties). This is due to
the fact that transitively self-similar Bianchi models arise as equilibrium
points in the dynamical state space of more general Bianchi models and
determine various stable and unstable submanifolds which may be identical to
some of the invariant sets of the orbits, consequently providing a way to
gain deeper insight into their asymptotic behavior \cite{Wainwright-Ellis}.
At the same time the assumption of self-similarity has the advantage of
reducing the Field Equations (FE) to a purely algebraic (closed) form which,
if not can be solved explicitly, is used for doing numerical simulations in
Bianchi cosmologies.

Up to now, the problem of finding the complete set of self-similar Bianchi
perfect fluid models with linear equation of state of the form $%
p=(\gamma-1)\mu$ is still open due to the (highly) non-linear character of
the resulting algebraic system. Considerable progress has been made for
vacuum and non-tilted perfect fluid models (orthogonal spatially homogeneous
cosmologies \cite{Ellis-MacCallum}) in which the whole class of self-similar
models has been determined \cite{Hsu-Wainwright}. In the case of tilted
perfect fluid models the situation is more complex and only some particular
results are known. For example Bradley \cite{Bradley} has stated that there
do not exist tilted dust self-similar models whereas it was proved recently
that irrotational Bianchi type VI$_0$ and models of type VII$_0 $, VIII and
IX are not self-similar \cite{Apostol-Tsampa7,Apostol9}.

The existent self-similar solutions have not been fully exploited apart from
the case of Bianchi type II models in which it has been proved \cite
{Hewitt-Bridson-Wainwright} that self-similar models may act as future
attractors for more generic models. In the general Bianchi type II
self-similar solution found by Hewitt \cite{Hewitt} the state parameter lies
in the interval $\gamma \in \left( \frac{10}7,2\right) $. This self-similar
model has zero vorticity and non-zero acceleration and was used to analyze
in full detail the whole class of type II models. In fact it was shown that $%
\gamma-$law perfect fluid Bianchi type II models, are future asymptotic to
the Collins-Stewart model \cite{Collins-Stewart1} when $\frac 23<\gamma <%
\frac{10}7$, consequently these models do not isotropise and the angle of
tilt becomes negligible at late times. At the value $\gamma =\frac{10}7$ the
tilt destabilizes the Collins-Stewart model and there is an exchange of
stability with the Hewitt model. In addition Hewitt solution has a four
dimensional stable manifold and play the role of the future attractor for
general Bianchi type II $\gamma-$law perfect fluid models. An interesting
feature of the stability analysis given in \cite{Hewitt-Bridson-Wainwright}
is that the equilibrium point $\gamma =\frac{14}9$ represents a second
bifurcation between the equilibria $\left( \frac{10}7,\frac{14}9\right) $
and $\left( \frac{14}9,2\right) $ (the latter does not correspond to a
self-similar model) and exhibits the property of the asymptotically extreme
tilt for models where the state parameter $\gamma$ belongs to the interval $%
\left( \frac{14}9,2\right) $.

Concerning the remaining Bianchi models the only known solution is due to
Rosquist and Jantzen (RJ) \cite{Rosquist1,Rosquist-Jantzen1} who have found
a family of Bianchi VI$_0$ models in which the state parameter $\gamma $
lies in the interval ($1.0411,1.7169$) ($\gamma \neq 10/9$). Although these
models are known over a decade it seems that their importance were not fully
understood. These models belong to the class $n_\alpha ^\alpha =0$, they are
rotating and consequently there are extra degrees of freedom adding to the
difficulty of qualitatively analysing general Bianchi VI$_0$ models.
Furthermore it was not known if this family represents the most general
self-similar tilted perfect fluid type VI$_0$ solution.

The overall goal of the present work is to fill this gap i.e. to show that
the self-similar Bianchi VI$_0$ tilted perfect fluid solution given in \cite
{Rosquist-Jantzen1} is \emph{the most general but the state parameter} $%
\gamma \in \left[ \frac 65,\frac 32\right) $. In addition, we present in
natural coordinates, the form of the self-similar metric, the Homothetic
Vector Field (HVF) and the kinematical and dynamical quantities of the model
and we discuss its physical implications in the asymptotic behavior of more
general Bianchi VI$_0$ models. In the rest of the paper we follow the
notations used in \cite{Apostol9}. However for convenience we present
briefly some of the basic results.

In \cite{Apostol9} the form of the transitively self-similar Bianchi VI$_0$
metric, the normalized tilted fluid velocity and the HVF are given in terms
of a set of arbitrary integration constants $c_{\alpha \beta },v_\alpha
,a,b,D,\psi $ (equations (3.29)-(3.31) of \cite{Apostol9}) where $\psi $ is
the homothetic factor. Setting $a=-(p_1+p_2)\psi +4\psi -b$ and $%
D=(2-p_2)\psi -b$ the \emph{frame }components of the self-similar metric and
the fluid velocity become: 
\begin{equation}
g_{\alpha \beta }=\left( 
\begin{array}{ccc}
c_{11}t^{2(p_1-1)} & c_{12}t^{p_1+p_2-2} & c_{13}t^{p_1} \\ 
c_{12}t^{p_1+p_2-2} & c_{22}t^{2(p_2-1)} & c_{23}t^{p_2} \\ 
c_{13}t^{p_1} & c_{23}t^{p_2} & c_{33}t^2
\end{array}
\right)  \label{sx1}
\end{equation}
\begin{equation}
\Delta _1=v_1t^{p_1-1}\qquad ,\Delta _2=v_2t^{p_2-1}\qquad ,\Delta _3=v_3t
\label{sx2}
\end{equation}
whereas the HVF assumes the form:

\begin{equation}
\mathbf{H}=\psi t\partial _t+\left[ (2-p_2)\psi-b\right] \partial _x+\left[
-(p_1+p_2)\psi +4\psi -b\right] y\partial _y+bz\partial _z.  \label{sx2_0}
\end{equation}
The FE and the Bianchi identities can be written: 
\begin{equation}
R_{ab}-\gamma \tilde{\mu}u_au_b-\frac{(2-\gamma )}2\tilde{\mu}g_{ab}=0
\label{sx2_1}
\end{equation}
\begin{equation}
\tilde{\mu}_{;a}u^a+\gamma \tilde{\mu}\tilde{\theta}=0  \label{sx2_2}
\end{equation}
\begin{equation}
(\gamma -1)\tilde{h}_a^k\tilde{\mu}_{;k}+\gamma \tilde{\mu}\dot{u}_a=0.
\label{sx2_3}
\end{equation}
In contrast with the existence of two hypersurface orthogonal Killing
Vectors (KVs) in type II models, self-similar Bianchi VI$_0$ models are
necessarily rotational and admit only one hypersurface orthogonal KV ($%
\mathbf{X}_1$ or $\mathbf{X}_3$)\cite{Hewitt}. Therefore it is convenient to
divide our analysis according to whether $X_1^kR_{k[a}X_{1b]}=0$ or $%
X_3^kR_{k[a}X_{3b]}=0$ (since the KVs $\mathbf{X}_1$,$\mathbf{X}_2$ form an
Abelian subgroup of the $G_3$ group of isometries the $\mathbf{X}_2$-case is
similar).

\underline{\textbf{Case }$X_1^kR_{k[a}X_{1b]}=0$}

We employ a new constant $s$ which is defined by the relation:

\begin{equation}
p_1+p_2=2\left( s+1\right) .  \label{sx2_4}
\end{equation}
Equation (\ref{sx2_2}) implies that the state parameter $\gamma $ is related
with $s$ via:

\begin{equation}
\gamma =\frac 2{2s+1}.  \label{sx2_5}
\end{equation}
From equation (\ref{sx2_1}) (or equivalently the existence of the
hypersurface orthogonal KV $\mathbf{X}_1$) it follows that $v_1=0$ and, in
order to avoid the orthogonal case, we assume $v_2,v_3\neq 0$. The resulting
system consists of a set of 10 highly non-linear algebraic equations (\ref
{sx2_1}) in 13 unknowns augmented by the system of equations (\ref{sx2_3}).

Imposing the physical restrictions: 
\begin{equation}
\tilde{\mu}>0,\quad \Gamma ^2>0,\quad \sigma ^2>0,\quad \omega ^2>0,\quad 
\dot{u}^a\dot{u}_a>0  \label{sx2a}
\end{equation}
we determine analytically the exact form of the self-similar metric and the
fluid velocity.\footnote{%
The majority of the computations have been made using the algebraic computer
package DERIVE.}

The rather lengthy and typical computations present no particular interest
and we summarize the results below (without loss of generality we set $%
c_{23}=-c_{11}=1$):

\begin{equation}
c_{12}=c_{13}=0  \label{sx3}
\end{equation}
\begin{eqnarray}
c_{33} &=&\left\{ \left( p_2-2\right) \left[ p_2+2\left( s-1\right) \right]
\right\} ^{-1}\left( p_2-s-1\right) [p_2^2+2p_2\left( 5s-2\right) - 
\nonumber \\
&&-4\left( 2s^2+3s-1\right) ]\times\left[ 3p_2-2\left( 2s+1\right) \right]
/[3p_2^4-p_2^3\left( s+14\right) +  \nonumber \\
&&+p_2^2\left( 23-10s^2\right) +2p_2\left( 4s^3+18s^2+3s-8\right) - 
\nonumber \\
&&-4\left( 2s+1\right) \left( 2s^2+3s-1\right) ]  \label{sx4}
\end{eqnarray}
\begin{eqnarray}
c_{22} &=&[3p_2^4-p_2^3\left( s+14\right) +p_2^2\left( 23-10s^2\right)
+2p_2\left( 4s^3+18s^2+3s-8\right) -  \nonumber \\
&&-4\left( 2s+1\right) \left( 2s^2+3s-1\right) ]/  \nonumber \\
&&\left\{[32s\left( p_2-s-1\right) ^3] \left( p_2-2\right)^{-1} \left[
p_2+2\left( s-1\right) \right]^{-1} \right\}  \label{sx5}
\end{eqnarray}

\begin{equation}
v_2=\frac{\left( 2-p_2\right) \left[ p_2+2\left( s-1\right) \right] }{%
4\left( p_2-s-1\right) }  \label{sx6}
\end{equation}
\begin{equation}
v_3=\frac{\left( p_2-s-1\right) \left[ 3p_2^2+2p_2\left( s-4\right) -4\left(
2s^2-s-1\right) \right] }{\left( 2-p_2\right) \left[ 3p_2^3-p_2^2\left(
s+8\right) -p_2\left( 10s^2+2s-7\right) +2\left( 4s^3+8s^2+s-1\right)
\right] }.  \label{sx7}
\end{equation}
The constant $p_2$ is related with the ''state parameter'' $s$ according to:

\begin{equation}
p_2=\frac{\sqrt{4s^2-36s+17}\left| 3s-1\right| -42s^2-17\left( s-1\right) }{%
17-36s}.  \label{sx8}
\end{equation}
The family of self-similar solutions (\ref{sx3})-(\ref{sx8}) has been given
previously in \cite{Rosquist-Jantzen1} using Hamiltonian methods and the
full group of automorphisms of the Lie algebra of isometries. The advantage
of the approach presented here is that we have used a natural choice for the
coordinates, adapted to the canonical 1-forms (\emph{metric approach}) hence
we can study the resulting models directly.

The kinematical and dynamical quantities of this family of self-similar
models are:

\begin{equation}
\tilde{\mu}=\frac{\left( 2s+1\right) \left( 2s+1-p_2\right) }{t^2}
\label{sx11}
\end{equation}
\begin{equation}
\dot{u}^a\dot{u}_a=\frac{\left( p_2-s-1\right) \left[ p_2+2\left( s-1\right)
\right] \left( 2s-1\right) ^2\left[ 3p_2-2\left( 2s+1\right) \right] }{%
t^2\left( p_2-2s-1\right) \left[ 3p_2-2\left( 3s+1\right) \right] }
\label{sx12}
\end{equation}
\begin{equation}
\Gamma ^2=\frac{3p_2^3-p_2^2\left( s+8\right) -p_2\left( 10s^2+2s-7\right)
+2\left( 4s^3+8s^2+s-1\right) }{\left( p_2-2s-1\right) \left[ 3p_2-2\left(
3s+1\right) \right] }  \label{sx13}
\end{equation}
\begin{eqnarray}
\sigma ^2 &=&\{3p_2^3\left( 56s^2-31s+8\right) -p_2^2\left(
56s^3+570s^2-249s+64\right) -  \nonumber \\
&&-4p_2\left( 140s^4-153s^3-161s^2+62s-14\right) +  \label{sx14} \\
&&+4\left( 112s^5+90s^4-179s^3-40s^2+21s-4\right) \}\times   \nonumber \\
&&\times \left\{ 6t^2\left[ 3p_2^2-p_2\left( 12s+5\right)
+12s^2+10s+2\right] \right\} ^{-1}  \nonumber
\end{eqnarray}
\begin{equation}
\omega ^2=\frac{s\left[ p_2+2\left( s-1\right) \right] ^2}{2t^2\left[
3p_2-2\left( 3s+1\right) \right] }.  \label{sx15}
\end{equation}
The positivity of the above quantities is ensured provided that: 
\begin{equation}
s\in \left( \frac 16,\frac 13\right) \Leftrightarrow \gamma \in \left( \frac
65,\frac 32\right) .  \label{sx16}
\end{equation}
It is easy to show that this solution belongs to the subclass of spatially
homogeneous models satisfying the constraint $n_\alpha ^\alpha =0$ hence it
represents a family of models which has a four dimensional stable manifold 
\cite{Hervik1}.

We also note that for the case where $n_\alpha ^\alpha \neq 0$ there is a
solution of the FE (\ref{sx2_1}) in which the constants $c_{\alpha \beta }$
are all non-vanishing. For these models the state parameter $\gamma $ takes
the values $\left( 1,\frac 65\right) \cup \left( \frac 32,2\right) $.
However none of these solutions is physically acceptable i.e. they do not
satisfy the inequalities (\ref{sx2a}).

\underline{\textbf{Case }$X_3^kR_{k[a}X_{3b]}=0$}

In this case $v_3=0$ and the solution of the FE implies that $p_1=p_2=\frac
43$ which by means of (\ref{sx2_4}) and (\ref{sx2_5}) the state parameter $%
\gamma =\frac 65$. The components of the self-similar metric and the fluid
velocity are:

\begin{equation}
c_{13}=c_{23}=1,\quad c_{22}=c_{11}  \label{sx17}
\end{equation}
\begin{equation}
c_{33}=\frac{9c_{11}^3+c_{11}^2\left( 324c_{12}^2-387c_{12}+40\right)
+c_{11}c_{12}\left( 387c_{12}-80\right) -c_{12}^2\left( 9c_{12}-40\right) }{%
12\left( c_{12}^2-c_{11}^2\right) \left[ c_{11}\left( 9c_{12}-1\right)
+c_{12}\right] }  \label{sx18}
\end{equation}
\begin{equation}
v_1=-v_2=-\frac{2\left[ c_{11}\left( 9c_{12}-1\right) +c_{12}\right] }{%
3\left( c_{11}+c_{12}\right) }  \label{sx19}
\end{equation}
\begin{equation}
9c_{11}^3\left( 54c_{12}-5\right) +c_{11}^2\left(
810c_{12}^2-387c_{12}+40\right) +c_{11}c_{12}\left( 441c_{12}-80\right)
-c_{12}^2\left( 9c_{12}-40\right) =0.  \label{sx20}
\end{equation}
We note that this model also belongs to the subclass $n^\alpha_\alpha=0$ and
admits the hypersurface orthogonal KV $\mathbf{X}_3$. Furthermore it was
recognized as an equilibrium point of the dynamical state space of the $%
\gamma-$law perfect fluid Bianchi VI$_0$ models \cite{Rosquist-Jantzen2}.

In conclusion we have shown that \emph{the general tilted Bianchi} VI$_0$%
\emph{\ self-similar solution is given by (\ref{sx3})-(\ref{sx8}) and (\ref
{sx17})-(\ref{sx20}) and the state parameter satisfies }$\frac 65\leq \gamma
<\frac 32$\emph{. }It should be noted that in \cite{Rosquist-Jantzen1} the
determination of the range of the state parameter $\gamma $ is based only on
the positivity of the energy density $\tilde{\mu}$. Indeed it is
straightforward to show that for $\gamma \in $ $\left( 1.0411,\frac
65\right) \cup \left( \frac 32,1.7169\right) $ ($\gamma \neq 10/9$) the
energy density is positive ($\tilde{\mu}>0$) but some (if not all) of the
kinematical quantities (\ref{sx12})-(\ref{sx15}) are negative.

The family of self-similar models found in the present paper could play a
similar role (as in the case of Bianchi type II models) in the asymptotic
behavior of general Bianchi type VI$_0$ models. In fact Barrow and Hervik 
\cite{Barrow-Hervik} have studied the (local) stability of tilted Bianchi
models at late times (in the neighborhood of the non-tilted equilibrium
point) and have shown that in type VI$_0$, the non-tilted Collins solution 
\cite{Collins2} (future attractor for nearby trajectories) is stable
whenever $\gamma \in \left( \frac 23,\frac 65\right) $. Therefore for $%
\gamma =\frac 65 $ the tilt destabilises the Collins model i.e. there is a
bifurcation which is associated with the stability change of the equilibrium
points (Collins model and the present solution). In addition for $\gamma \in
\left( \frac 65,\frac 32\right)$ the present family of models is
asymptotically tilted ($v_3\neq 0$) and has a four dimensional stable
manifold \cite{Hervik1} hence it is possible to play the role of the future
attractor at least for the subclass of tilted Bianchi VI$_0$ models
satisfying $n_\alpha ^\alpha =0$ (we recall that for general Bianchi VI$_0$
models i.e. when $n_\alpha ^\alpha \neq 0$ the dynamical state space is
seven dimensional).

In view of the above results and the similarities between types II and VI$_0$
one may conjecture that the subclass $n_\alpha ^\alpha =0$ of Bianchi type VI%
$_0$ models are possible to be asymptotically tilted at late times for $%
\gamma \in \left( \frac 65,2\right) $. Furthermore a preliminary analysis on
this class of tilted Bianchi VI$_0$ models indicate that whenever $\gamma
\in \left( \frac 32,2\right) $, the tilt is becoming extreme at late times 
\cite{Hervik2}.\vspace{0.4cm}

The author is grateful to Sigbj\o rn Hervik for enlightening discussions and
comments on an earlier draft of the manuscript and for bringing into the
author's attention some of the results of his work.

\end{document}